\documentclass[aps,reprint,amssymb,superscriptaddress]{revtex4-2}  
\usepackage{times,units,braket}
\usepackage{graphicx}  
\usepackage{graphics}
\usepackage{dcolumn}   
\usepackage{bm}        
\usepackage{amssymb}   
\usepackage{amsmath,upgreek}
\usepackage{natbib}
\usepackage[colorlinks,citecolor=blue]{hyperref}
\usepackage{color}
\usepackage{amssymb}
\usepackage{changepage}
\usepackage{float}
\usepackage{caption}
\begin{document}

\widetext

\title{Observation of relativistic Bohmian dynamics}
\author{Yun-Fei Wang}
\author{Hui Wang}
\email{whui@ustc.edu.cn}
\author{Tong Zhang}
\author{Yi-Teng Ye}
\author{Xiao-Yu Wang}
\author{Ming-Cheng Chen}
\affiliation{%
    Hefei National Laboratory for Physical Sciences at the Microscale and Department of Modern Physics, University of Science and Technology of China, Hefei 230026, China %
}
\affiliation{%
    Shanghai Branch, CAS Center for Excellence in Quantum Information and Quantum Physics, University of Science and Technology of China, Shanghai 201315, China%
}
\affiliation{%
    New Cornerstone Science Laboratory, Shanghai Research Center for Quantum Sciences, Shanghai 201315, China%
}
\author{Chao-Yang Lu}
\email{cylu@ustc.edu.cn}
\author{Jian-Wei Pan}
\email{pan@ustc.edu.cn}
\affiliation{%
    Hefei National Laboratory for Physical Sciences at the Microscale and Department of Modern Physics, University of Science and Technology of China, Hefei 230026, China %
}
\affiliation{%
    Shanghai Branch, CAS Center for Excellence in Quantum Information and Quantum Physics, University of Science and Technology of China, Shanghai 201315, China%
}
\affiliation{%
    New Cornerstone Science Laboratory, Shanghai Research Center for Quantum Sciences, Shanghai 201315, China%
}

\date{\today}

\begin{abstract}
Bohmian mechanics, also referred to as the de Broglie–Bohm pilot-wave theory, represents a deterministic and nonlocal interpretation of quantum mechanics \cite{De27,Bohm52,Bohm52b}. Central to this framework is a description of quantum motion in terms of particle trajectories, whose reconciliation with relativity remains a fundamental challenge. Although relativistic guidance laws have been proposed, their predicted dynamics has remained experimentally untested. Here we use weak measurements \cite{Aharonov88,wiseman07,Lundeen11,Dressel14} in a single-photon interferometer to jointly reconstruct the photon energy and momentum weak values and the associated relativistic velocity field. The reconstructed average trajectories agree with theoretical predictions. Near destructive interference, the flow exhibits local reconstructed velocities exceeding the speed of light and negative values of the effective squared-mass parameter. Our results establish an experimental basis for investigating the physical meaning of quantum trajectories and their role in describing motion in relativistic spacetime.
\end{abstract}

\maketitle

While quantum mechanics is spectacularly successful in predicting results of measurements at the microscopic level, fundamental debates persist regarding the quantum measurement problem \cite{Sch05,Leggett05,Bassi13} and the nature of physical reality \cite{Bohr36}. These unresolved issues have stimulated various quantum-mechanical interpretations \cite{Omnes18,Sak95,Everett57,De27,Bohm52,Bohm52b,Bohm62,de1987interpretation}, among which Bohmian mechanics is a nonlocal hidden-variable framework \cite{Bell82}. Unlike the orthodox Copenhagen interpretation that relies solely on the wavefunction, Bohmian mechanics emphasizes a dual description $(\psi, Q_i)$ involving both the wavefunction $\psi$ and the configuration $Q_i$ (or trajectories). Here, particles follow deterministic trajectories guided by a ``pilot wave" that evolves according to the Schrödinger equation \cite{Bohm52,Bohm62,de1987interpretation}.

Theoretically, Bohmian mechanics has the capacity to recover predictions of standard non-relativistic quantum mechanics \cite{Bohm52b,Hirschfelder74,Philippidis79,Dewdney88,Durr92,Allori08,Durr09}. More importantly, it has evolved from a conceptual construct into an experimentally testable framework: the integration of weak measurement techniques \cite{Aharonov88,wiseman07,Lundeen11,Dressel14} has enabled the experimental observation of Bohmian trajectories in the non-relativistic regime \cite{Kocsis11,Mahler16,Frumkin22}. 

\begin{figure*}
    \centering
    \includegraphics[width=0.85\linewidth]{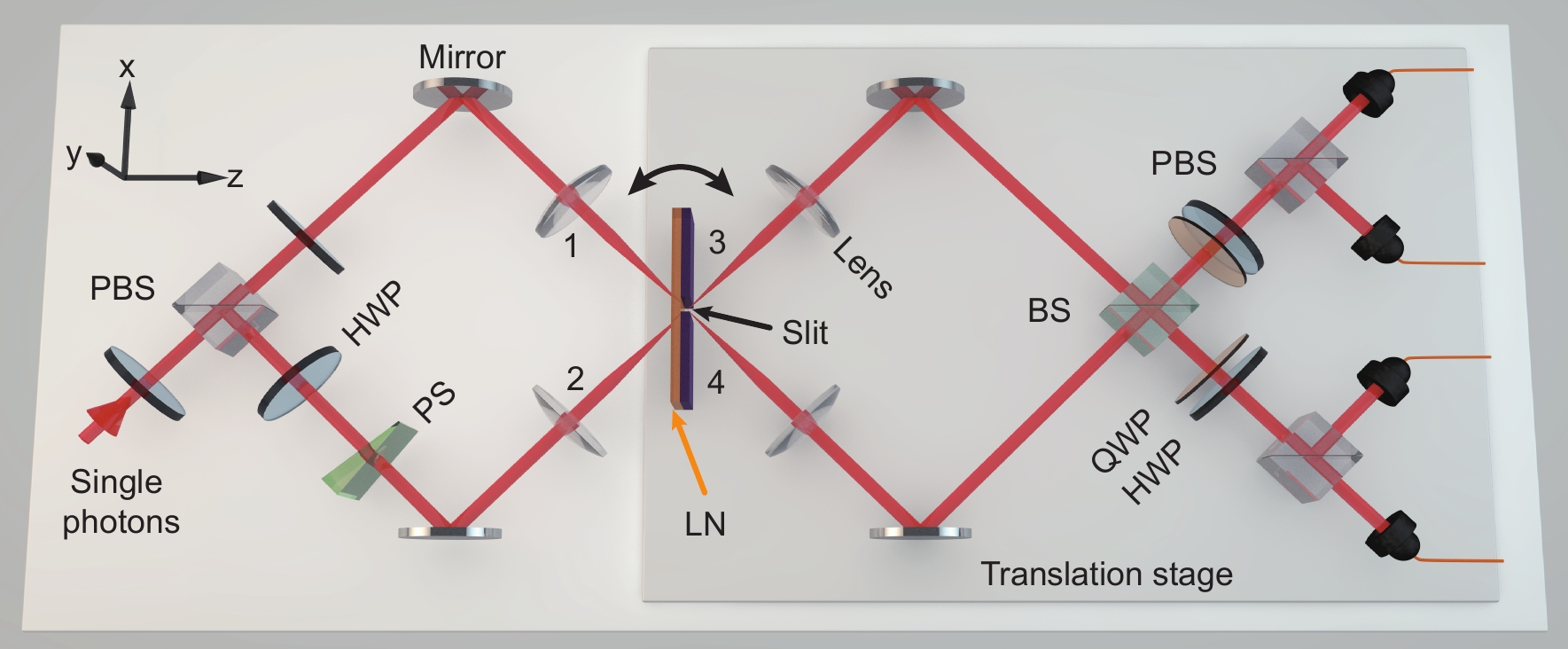}
    \caption{\textbf{Experimental setup.} An illustration of the experimental setup. The single photons emitted from an InAs/GaAs quantum dot confined in a microcavity are directed into an interferometer. A weak measurement is conducted on the photon polarization state $\ket{+}$ prepared by a polarizing beam splitter (PBS) and two half-wave plates (HWPs), with a birefringent LN plate featuring a 1 $\mu$m-wide slit on its surface positioned at the focal plane of four lenses. Another phase shifter (PS) on a movable precision translation stage placed in one path is used to calibrate the relative phase during measurement. In order to intricately scan the interference fringes, all optical components in the shaded region are placed on a movable precision translation stage. The slit and the two lenses on stage participate in forming a post-selection and imaging system. The two beams are recombined by a beam splitter (BS). We only use the photon count of one path after the BS to avoid the systematic calibration error. Finally, the photon states are analyzed in the polarization basis of $(|H\rangle,\ |V\rangle)$ using two superconducting nanowire single-photon detectors. 
    }
    \label{fig1}
\end{figure*}

Despite these successes, reconciling Bohmian mechanics with relativity remains a significant challenge. The Schrödinger equation is not Lorentz covariant, and the nonlocal dependence of Bohmian dynamics on the configuration of an entangled system complicates its reconciliation with relativistic spacetime structure \cite{Berndl96,Durr14}. Although numerous attempts have been made to extend this framework into the relativistic domain \cite{Bohm06,Berndl96,Horton02,Nik04,Nik09,Dewd10,Niko10,Durr14,Foo22,Bloch23,Foo23}, experimental evidence substantiating relativistic effects within a Bohmian context remains elusive.
 
Here we present an experimental study of relativistic Bohmian dynamics based on an operational weak-value formulation proposed in Ref.~\cite{Foo22}. In this framework, ensemble-averaged relativistic Bohmian trajectories of single photons can be reconstructed from a velocity field defined by the ratio of the weak values of momentum and energy (in units where $\hbar=c=1$):
\begin{equation}
\mathbf{v}(x,z)=\frac{\langle \hat{\mathbf{p}}_w \rangle (x,z)}{\langle \hat{E}_w \rangle (x,z)}.
\label{velocity}
\end{equation}
Throughout this work, $\langle \hat{A}_w\rangle$ denotes the real part of the photon-count-weighted weak value of an operator $A$. It has been theoretically shown that this velocity field satisfies both the relativistic velocity addition rule and continuity equation given by scalar Klein-Gordon (KG) equation. 

The same weak-value formalism also provides an effective squared mass, originally introduced by de Broglie \cite{de1987interpretation} and extended to weak values in Ref.~\cite{Bloch23}:
\begin{equation}
\bar{m}_{\text{eff}}^2=(\langle \hat{E}_w \rangle)^2-|\langle \hat{\mathbf{p}}_w \rangle|^2.
\label{mass}
\end{equation}
This quantity provides a local classification of the reconstructed motion. Positive, zero and negative values correspond to timelike, lightlike and spacelike motion, respectively. Accordingly, a negative effective squared mass and a superluminal reconstructed velocity are two signatures of the same local kinematic property. Moreover, for the massless and monochromatic scalar-field, the effective squared mass parameter coincides with the quantum potential, which plays a central role in the photon dynamics described by this formalism (Supplementary Section 3.2).

Relativistic Bohmian theories predict a local flow field with distinctive relativistic kinematics, but whether this field can be reconstructed and tested experimentally remains unresolved. What kind of motion can the relativistic trajectories reconstructed via weak measurements really be understood to be? Here, we test whether trajectories reconstructed from these measured weak values reproduce the theoretical predictions for relativistic Bohmian flow, and whether the same local fields exhibit the predicted superluminal reconstructed velocities and negative effective squared mass. Addressing these questions would establish an experimental connection between the kinematic content of relativistic Bohmian dynamics and measurable quantum observables, providing a basis for testing and developing relativistic theories of quantum motion. 
\vspace{\baselineskip}

\noindent{\textbf{Weak measurement}}

To this end, a double-slit interferometer is employed to study the relativistic Bohmian mechanics (Fig.~\ref{fig1}), where the photon polarization serves as a pointer to perform joint weak measurements of $\langle \hat{\mathbf{p}}_w \rangle$ and $\langle \hat{E}_w \rangle$. Single photons are generated via a self-assembled InAs/GaAs quantum dot (QD) embedded in a tunable polarized microcavity \cite{wang19,ding24}. The measured second-order correlation function $g^2(0)=0.04(1)$ confirms the high-purity single-photon emission (Supplementary Section 4). 

Initially, single photons in both paths are prepared in the linearly polarized state $\ket+=\frac1{\sqrt2}(\ket{H}+\ket{V})$ by using two half-wave plates (HWPs). Subsequently, two beams are focused to a spot size of 8.3 ± 0.3 $\mu$m using a pair of lenses ($f = 10$ mm) and spatially overlapped. The region of interest is an $18 \times 18\ \mu \text{m}^2$ overlapping area. The incident angles of the two paths are ± $45^\circ$. This means that the velocity components $|v_x |$ and $v_z$ are approximately $\frac{\sqrt{2}}{2}c$ when only one arm of the light is illuminated and $c$ denotes the speed of light.

The key component for performing weak measurements is a birefringent plate, comprising a 500 nm-thick $x$-cut lithium niobate (LN) thin film on a 500 $\mu$m SiO$_2$ substrate. A 1 $\mu$m-wide slit is affixed to the LN surface for the purpose of scanning the light beam (Supplementary Section 5). When single photons pass through the LN plate, positioned in the confocal plane of the lens, their initial polarization state $\ket{+}$ undergoes a small change of $\delta\varphi$ on the Bloch sphere, given $\varphi=\pi/2+\delta \varphi$ and $\ket{\psi}=\cos(\varphi/2)\ket{H}+\sin(\varphi/2)e^{i\phi}\ket{V}$, where $\phi$ represents the relative phase between $\ket{H}$ and $\ket{V}$.

Critically, the polar angle $\varphi$ of the polarization pointer is coupled to a linear combination of the photon momentum and energy weak values. The calibrated responses of the five LN configurations therefore provide complementary constraints on the same underlying complex field (Supplementary Section 7). We jointly reconstruct this field from the raw H and V photon counts using a calibrated optical response and a Poisson likelihood. The photon-count-weighted energy and momentum weak values are then calculated from the fitted field representation (Supplementary Section 8).

\begin{figure}
    \centering
    \includegraphics[width=89mm]{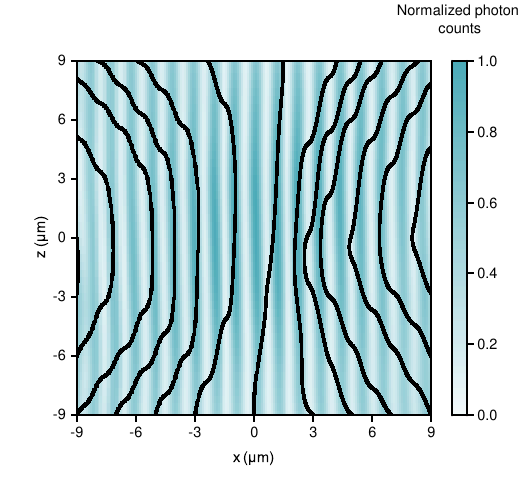}
    \caption{\textbf{Reconstructed average Bohmian trajectories in the relativistic domain.} The trajectories (depicted as black lines) were reconstructed across a range of $\pm$9 $\mu$m along both $x$ and $z$ axes. The cyan map illustrates the measured interference fringes obtained by shifting the slit on the LN plate with step sizes of 100 nm and 400 nm along the $x$ and $z$ axes, respectively. The trajectory bending within the regions exhibiting destructive interference could be observed. 
    }
    \label{fig2}
\end{figure}

To acquire the angle $\varphi$, two beams are recombined in a beam splitter (BS), and the polarization state at one of the ports is measured in the polarization basis of $(|H\rangle,\ |V\rangle)$. In this case, $\varphi$ can be determined by
\begin{equation}
\varphi=2 \arctan \sqrt{\frac{I_V}{I_H}},
\label{weak}
\end{equation}
where $I_H$ and $I_V$ denote the photon numbers in the $H$ and $V$ polarization states, respectively. Further technical details on the calibration of the coefficients are provided in the Methods and Extended Data Fig.~\ref{ED1}.
\vspace{\baselineskip}

\noindent{\textbf{Average photon trajectories}}

To obtain the average photon trajectories, a precise, step-by-step mapping of the interference fringes alongside the systematic recording of weak momentum and energy values at each position is required. For this purpose, all elements in the shaded region in Fig.~\ref{fig1} are placed on a precision translation stage. We employ scanning steps of 100 nm along the $x$-axis and 400 nm along the $z$-axis, covering a total range of 18 $\mu$m within the region of maximum intensity. This high-resolution mapping comprises 8,326 individual data points. At each step, we utilize a 3-second integration time for $I_H$ and $I_V$, with an additional 2-second interval that allows for stage translation and stabilization. The measured interference map obtained from the sum of $I_H$ and $I_V$ is shown in cyan in Fig.~\ref{fig2}.

\begin{figure*}
    \centering
    \includegraphics[width=183mm]{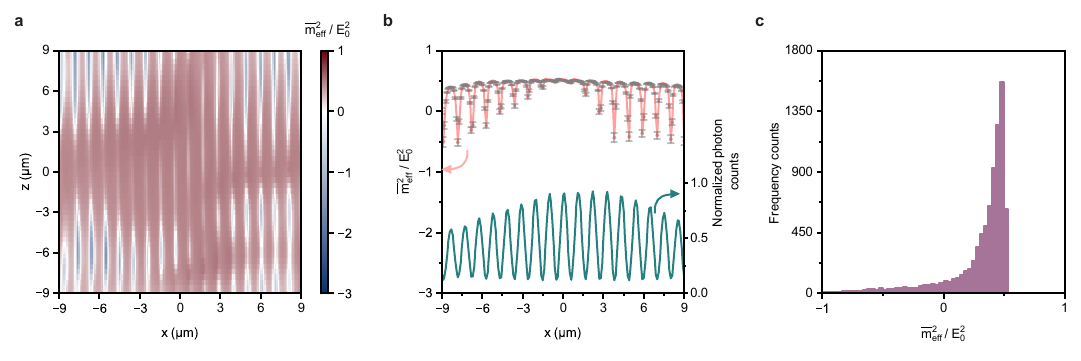}
    \caption{\textbf{Effective squared photon mass}. \textbf{a}, The map of determined effective squared mass $\bar{m}_{\text{eff}}^2$ over a range of $\pm$9 $\mu$m along both $x$ and $z$ axes. Note that $\bar{m}_{\text{eff}}^2$ is normalized by the squared photon energy. In the constructive regions, $0<\bar{m}_{\text{eff}}^2<1$, while in the destructive regions, $\bar{m}_{\text{eff}}^2<0$. As the determined $\bar{m}_{\text{eff}}^2$ corresponds to the quantum potential, this is also an illustration of quantum potential for interference fringes. \textbf{b}, An example of $\bar{m}_{\text{eff}}^2$ (red) with statistical error bars (gray) and the corresponding interference fringe (cyan) at $z=4\ \mu m$. \textbf{c}, The statistical distribution of $\bar{m}_{\text{eff}}^2$ depicted in \textbf{a}.
    }
    \label{fig3}
\end{figure*}

At each spatial position, a single measurement configuration provides one polarization readout that depends on the three weak value $\langle p_{x,w}\rangle$, $\langle p_{z,w}\rangle$ and $\langle E_w\rangle$. To address this challenge, we replicated the process with five distinct measurements, using two LN plates with optical axes perpendicular and parallel to the $y$-axis, positioned at angles of $-10^{\circ}$, $0^{\circ}$ and $10^{\circ}$ relative to the $x$ axis, respectively. Under the calibrated linear-response approximation, these configurations probe five combinations of the three weak values at each spatial position (Supplementary Section 7.3). 

We jointly fit the raw $H$ and $V$ photon counts across all five configurations and spatial sampling positions using the common field representation described in Supplementary Section 8. The forward model incorporates the calibrated LN response and scan-row phase variations, with photon-counting noise described by a Poisson likelihood. The independently characterized quantum-dot linewidth sets the frequency scale of the Lorentzian soft constraint used in the fit. The fitted representation directly yields the photon-count-weighted energy and momentum weak values integrated over the detector gate (Supplementary Section 8.4). These quantities are evaluated at the 8,326 spatial sampling positions and shown in Extended Data Fig.~\ref{ek}. Their ratio defines the velocity field v(x, z) through Eq.~\ref{velocity}, with its components shown in Extended Data Figs. 3a and 3b. Finally, by employing the fourth-order Runge–Kutta method, the average relativistic Bohmian trajectories are reconstructed from the measured velocity field. Despite experimental noise sources---such as phase instability and detector dark counts---the reconstructed average trajectories remain consistent with theoretical simulations (Extended Data Fig.~\ref{intensity_simulation}, Supplementary Section 9)

It is important to emphasize that the reconstructed trajectories in Fig.~\ref{fig2} do not imply that individual photons follow these specific paths. Rather, they represent an ensemble-averaged behavior: at each spatial coordinate $(x, z)$, the measured velocity $\mathbf{v}(x, z)$ reflects the average velocity of all photons passing through that point. Furthermore, a pronounced bending of the trajectories is observed in regions of destructive interference. This curvature is related to local velocities that exceed the speed of light (Extended Data Figs.~\ref{vel}c and \ref{vel}d). Crucially, this phenomenon does not violate the principles of relativity. Since $\mathbf{v}(x, z)$ is a local flow velocity rather than a signal-carrying velocity, it does not allow superluminal information transfer. 
\vspace{\baselineskip}

\noindent{\textbf{Effective squared mass}}

Subsequently, the effective squared-mass parameter, defined in Eq.~\ref{mass}, is determined from the measured weak values of the photon energy and momentum, as shown in Fig.~\ref{fig3}a. All values are normalized by $E_0^2$, where $E_0$ denotes the central photon energy in the same $\hbar=c=1$ units. The reconstructed map reveals a structured spatial pattern across the interference region.

In constructive-interference regions, $\bar{m}_{\mathrm{eff}}^2$ remains positive and is bounded by $E_0^2$. As the field approaches destructive-interference regions, the parameter reverses sign and develops pronounced negative excursions. Fig.~\ref{fig3}b shows this behavior along $z=4\ \mu m$, where the negative regions of $\bar{m}_{\mathrm{eff}}^2$ are spatially associated with the minima of the interference pattern. These regions also correspond to the strongest bending of the reconstructed trajectories and to regions where the reconstructed velocity exceeds $c$ (Extended data Fig.~\ref{vel}d). Thus, in the perspective of relativistic Bohmian mechanics, the interference pattern does not merely modulate the detected photon intensity; it organizes the reconstructed flow into distinct timelike-like and spacelike-like domains.

For the massless scalar-field description used here, the same spatial structure can approximately be interpreted as a spatially resolved quantum-potential landscape (Supplementary Section 6). In Bohmian mechanics, this potential arises from the curvature of the wave amplitude, i.e. the $\partial^\mu \partial_\mu R/R$ term, rather than from an external interaction. The observed sign-changing pattern therefore reflects the sign of the quantum potential, which controls whether the local four-momentum of the massless Bohmian particle is timelike, null, or spacelike. This changes the local on-shell condition and hence the reconstructed flow, without implying any change in the photon’s physical rest mass.

The distribution shown in Fig.~\ref{fig3}c summarizes the measured values and confirms that all the positive effective squared-mass parameter remains bounded by $E_0^2$, whereas negative values occur in localized destructive-interference regions. These negative values characterize the reconstructed weak-value flow and do not imply an imaginary photon rest mass. The physical rest mass of the photon always remains zero.
\vspace{\baselineskip}

\noindent{\textbf{Discussion and outlook}}

In summary, our joint weak-value reconstruction provides an operational test of relativistic Bohmian flow in single-photon interference. The measured flow reproduces the predicted average trajectories and develops localized regions with reconstructed velocities exceeding $c$ and negative values of $\bar{m}_{\mathrm{eff}}^2$ near destructive interference. Their spatial correlation shows that interference reshapes the local on-shell condition of the reconstructed field, rather than merely redistributing photon intensity. In the massless scalar-field description, this structure provides an operational view of the associated quantum-potential landscape. Our results link the kinematic structure of relativistic Bohmian dynamics to experimentally reconstructed weak-value fields, providing a benchmark for further theoretical development and testing. More broadly, our work opens an experimental route to reassessing the physical meaning of quantum trajectories and their role in describing motion in relativistic spacetime.

Our reconstruction yields a nonzero difference between the normalized divergences of the number and energy-weighted currents (Supplementary Section 10). However, the present uncertainties and reconstruction variability prevent a conclusive assessment of which candidate density better describes locally conserved transport under the reconstructed flow. Resolving which conserved quantity underlies this flow is therefore central to the physical interpretation of experimentally reconstructed relativistic Bohmian trajectories.

These quantitative observations also reveal a fundamental conceptual tension for point-particle Bohmian mechanics in flat spacetime. This tension arises if the velocity field reconstructed from weak measurements is identified with the physical velocity of a point particle. Although the superluminal flow does not enable superluminal information transfer, it remains difficult to reconcile with strict relativistic causality for point-like entities. Overcoming this while preserving relativistic causality and deterministic dynamics may require a broader field-theoretic or curved-spacetime formulation \cite{Bohm52b,Foo22,wang2026}.

In the near future, relativistic Bohmian mechanics for correlated photons can be realized based on our experiment \cite{Foo23}. Moreover, this approach can be extended to the Dirac equation for fermions to unveil more intriguing physics. Bohmian mechanics, whether or not it brings us closer to the essence of nature, offers us an unparalleled perspective, not found within existing theoretical frameworks, to comprehend nature. We hope our work can inspire more investigations on relativistic Bohmian mechanics to elucidate a broad range of quantum phenomena.
\vspace{\baselineskip}

\begin{acknowledgments}
We thank Timothy Ralph and Dao-Neng Gao for very helpful discussions. This work is supported by the National Natural Science Foundation of China (T2422024, 12204460), the National Key R\&D Program of China (2019YFA0308700), the Chinese Academy of Sciences, the Anhui Initiative in Quantum Information Technologies (AHY060000), the Science and Technology Commission of Shanghai Municipality (2019SHZDZX01), the Innovation Program for Quantum Science and Technology (2021ZD0301400), and the USTC Research Funds of the Double First-Class Initiative (YD9990002020).
\end{acknowledgments}

\bibliographystyle{naturemag}
\bibliography{references}

\clearpage

\noindent{\large\textbf{Methods}}
\\

\noindent{\textbf{KG Bohmian velocity = KG flow velocity}}

We next clarify the definition of the velocity field used in this work and establish its formal connection to the flow velocity associated with the KG equation.

In post-selection of position $\mathbf{x}$, weak momentum and weak energy are
\begin{equation}
\mathbf{p}_w = \operatorname{Re}\frac{\langle \mathbf{x}|\hat{\mathbf{p}}|\psi\rangle}{\langle \mathbf{x}|\psi\rangle},
\qquad 
E_w = \operatorname{Re}\frac{\langle \mathbf{x}|\hat{E}|\psi\rangle}{\langle \mathbf{x}|\psi\rangle}.
\label{wv}
\end{equation}
For a single polarized optical-field envelope, the wavefunction can be expressed as
\begin{equation}
\psi(\mathbf{x},t)=R(\mathbf{x},t)e^{iS(\mathbf{x},t)},
\label{wf}
\end{equation}
where $R(\mathbf{x},t)$ and $S(\mathbf{x},t)$ denote the real amplitude and the phase function, respectively. Substituting the operators $\hat{\mathbf{p}}=-i\nabla$ and $\hat{E}=i\partial_t$ along with Eq.~\ref{wf} into Eq.~\ref{wv} yields
\begin{equation}
\mathbf{p}_w = \nabla S, 
\qquad 
E_w = -\partial_t S.
\end{equation}
Consequently, the Bohmian velocity field $\mathbf{v}$ takes the form
\begin{equation}
\mathbf{v} = \frac{\mathbf{p}_w}{E_w} = \frac{\nabla S}{-\partial_t S}.
\label{velocity_weak}
\end{equation}

The flow velocity given by the KG equation is
\begin{equation}
\mathbf{v}_{\text{flow}}=\frac{\mathbf{j}_K}{\rho_K},
\label{KG-vel}
\end{equation}
where $\mathbf{j}_K=\text{Im}\psi^*\nabla\psi$ and $\rho_K=-\text{Im}\psi^*\partial_t\psi$. Substituting $\psi=Re^{iS}$ into Eq.~\ref{KG-vel} gives
\begin{equation}
\mathbf{v}_{\text{flow}}=\frac{\mathbf{j}_K}{\rho_K}=\frac{\nabla S}{-\partial_t S}.
\label{KG_continuity}
\end{equation}

By comparing Eq.~\ref{velocity_weak} and Eq.~\ref{KG_continuity}, we demonstrate that the velocity field extracted from weak measurements is mathematically equivalent to the local ensemble flow velocity given by the KG equation. Crucially, this quantity reflects a local ensemble average rather than the physical velocity of an individual photon under strong projective measurements. Furthermore, this local flow velocity is distinct from the signal velocity. Therefore, even if the measured velocity field comfortably exceeds the speed of light in regions of destructive interference, this phenomenon does not imply superluminal information transmission.\\

\noindent{\textbf{Determination of the coefficients (a, b, c)}}

We employ another independent optical setup (Extended Data Fig.~\ref{ED1}a), where the optical axis of the LN plate is perpendicular or parallel to the $y$-axis. It should be noted that the two types of LN plate usage are motivated by the requirement to obtain more than three distinct sets of parameters while ensuring alignment with the measurement configuration illustrated in Fig.~\ref{fig1}. We find that the polar angle $\varphi$ on the Bloch sphere depends approximately linearly on the momentum and energy in the regime of our experiment. In this case, the relation is $\varphi=a_{i,x}\braket{p_{x, w}} + a_{i,z}\braket{p_{z,w}} + b_i \braket{E_w} + c_i$ ($i$ = 1,2), which can be noted as $\varphi=a_i \braket{p_{\perp,w}}+b_i \braket{E_w}+c_i$ due to the characteristic of the $x$-cut LN (Supplementary Section 7), where $p_{\bot}$ denotes the momentum component that is perpendicular to the surface of the LN plate.

By scanning both $p_{\bot}$ and $E$ of incident light and recording the corresponding polar angle $\varphi$, we can determine all coefficients ($a_i, b_i, c_i$) by linear fitting. To this end, the LN plate with the optical axis perpendicular (parallel) to the $y$-axis, together with the downstream optical elements, is rotated over the range of $(-1.5^{\circ},1.5^{\circ})$ to slightly change the momentum $p_{\bot}$. The photon energy $E$ is equivalently varied by scanning the laser's wavelength from 1529.83 to 1564.95 nm. Similar measurements are repeated three times with angles of $35^{\circ}$, $45^{\circ}$ and $55^{\circ}$ between the LN plate and the incident light beam, and all measured values of $\varphi$ are listed in Extended Data Figs.~\ref{ED1}b and~\ref{ED1}c. From linear fits to the data, the two sets of parameters ($a_i, b_i, c_i$) for LN plates with the optical axis perpendicular and parallel to the $y$-axis are $(2.04\times10^{-7}\ \text{rad}\cdot\text{m}, -2.75\times10^{-15}\ \text{rad}\cdot\text{s}, 4.08\ \text{rad})$ and $(1.64\times10^{-7}\ \text{rad}\cdot\text{m}, -2.71\times10^{-15}\ \text{rad}\cdot\text{s}, 4.21\ \text{rad})$, respectively. The fitted variables $p$ and $E$ are the optical-unit quantities $p/\hbar$ and $E/\hbar$, so the original experimental coefficient units remain $\text{rad}\cdot\text{m}$ and $\text{rad}\cdot\text{s}$.
\\

\setcounter{figure}{0}
\captionsetup[figure]{name=\textbf{Extended Data Fig.}}

\begin{figure*}
    \centering
    \includegraphics[width=183 mm]{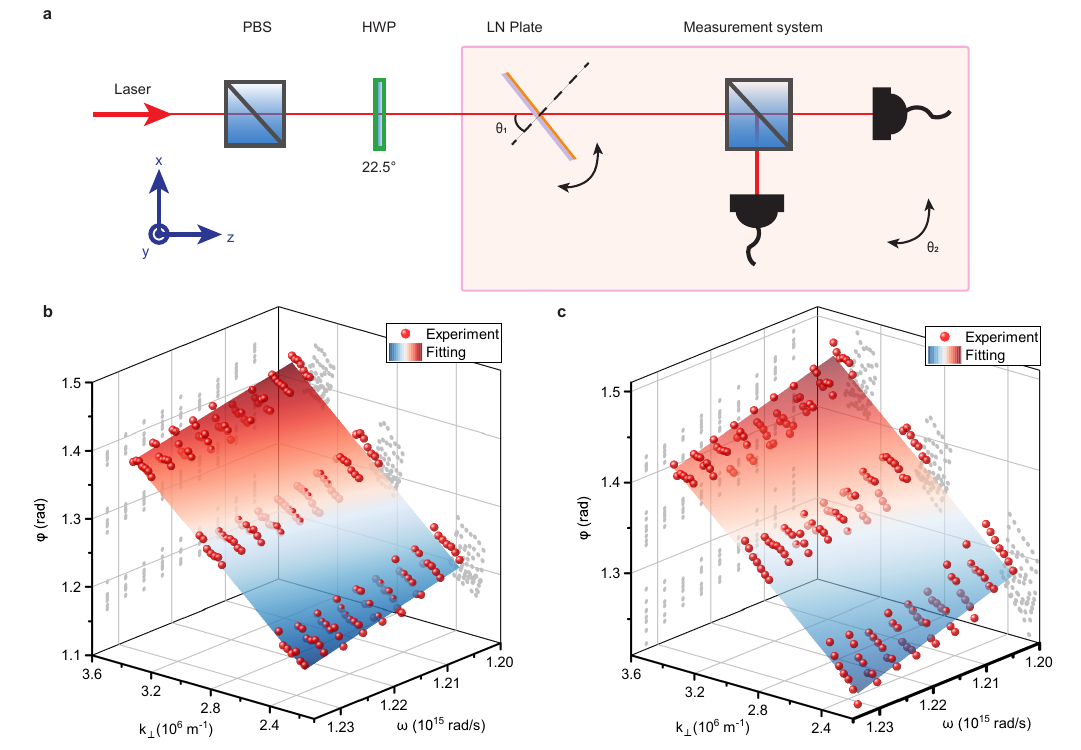}
    \caption{\textbf{Determining coefficients}. 
    \textbf{a}, Optical setup to determine the relation between the polar angle $\varphi$ of the polarization state and the momentum and the energy of the photons. A laser with a $\sim$kHz linewidth is prepared in the $|+\rangle$-polarized state using a PBS and an HWP set at 22.5$^{\circ}$. Then the measurement system is rotated by $\theta_2$ and the laser wavelength is varied to tune the photon energy. The resulting polar angle $\varphi$ of the polarization state is determined from the light intensities at the two output ports of the PBS.
    \textbf{b} and \textbf{c} represent linear fitting between the polar angle $\varphi$ and the momentum $p_{\perp}$ and energy $E$ (red dots) obtained when the optical axes of two LN plates are aligned along the $x$ and $y$ axes, respectively. The colored areas symbolize the theoretical fitting using a linear function, while the gray dots represent the projection of $\varphi$ values. These results indicate that $\varphi$ is indeed a nearly linear function of the momentum $p_{\perp}$ and energy $E$.
    }
    \label{ED1}
\end{figure*}

\newpage

\begin{figure*}
    \centering
    \includegraphics[width=183 mm]{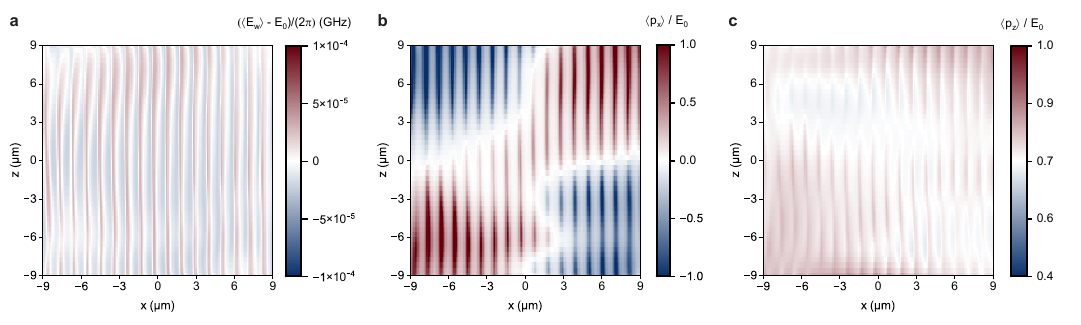}
    \caption{\textbf{Weak value distribution of $\langle\hat{E}_w\rangle$, $\langle\hat{p}_{x,w}\rangle$, and $\langle\hat{p}_{z,w}\rangle$}. \textbf{a}, Measured distribution of $\langle\hat{E}_w\rangle$. The energy of a single photon at 1550 nm is near $1.2 \times 10^{15}$ rad/s, but the dispersion of the energy weak value is observed. \textbf{b} and \textbf{c} show the measured distributions of $\langle\hat{p}_{x,w}\rangle$ and $\langle\hat{p}_{z,w}\rangle$, respectively. $\langle\hat{p}_{x,w}\rangle$ takes both positive and negative values: the positive values are mainly distributed in the diagonal region, while the negative values are primarily located in the anti-diagonal region. This directly explains why the measured trajectory deviates in the $x$-direction. The determined $\langle\hat{p}_{z,w}\rangle$ takes all positive values.
    }
    \label{ek}
\end{figure*}

\clearpage

\begin{figure*}
    \centering
    \includegraphics[width=183mm]{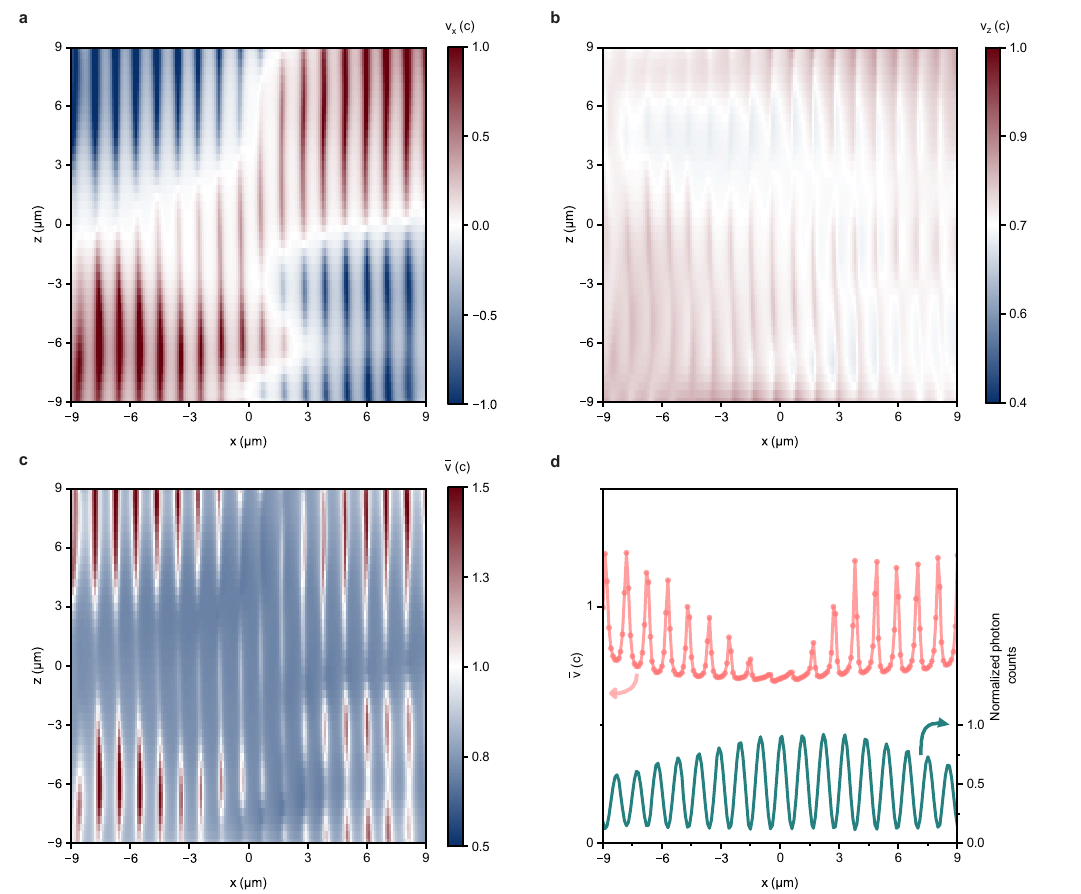}
    \caption{\textbf{The determined velocity field}. \textbf{a} and \textbf{b} show the measured $v_x$ and $v_z$, respectively, normalized by the speed of light $c$. $v_x$ takes both positive and negative values, which is mainly determined by $\langle\hat{p}_{x,w}\rangle$, whereas $v_z$ is always positive. \textbf{c} presents the distribution of the velocity magnitude $|v|=\sqrt{v_x^2 + v_z^2}$. In the destructive interference regions, $|v|$ exceeds $c$. In the constructive interference regions, $|v|$ is approximately $c\cos45^{\circ}$, as illustrated in \textbf{d}. There, the cyan line represents the measured interference fringe at $z=4\ \mu m$, and the red line shows the corresponding $|v|$. 
    }
    \label{vel}
\end{figure*}

\begin{figure}
    \centering
    \includegraphics[width=89mm]{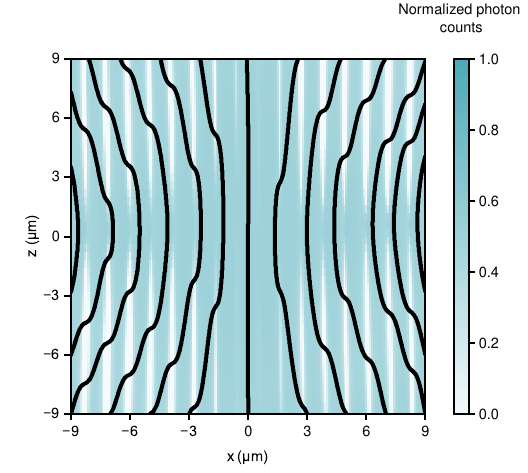}
    \caption{\textbf{Simulated average photon trajectories}. The cyan colormap illustrates the simulated photon interference fringes. The black lines denote the simulated average trajectories. 
    }
    \label{intensity_simulation}
\end{figure}

\end{document}